# Solid lubrication by wet-transferred solution-processed graphene flakes: dissipation mechanisms and superlubricity in mesoscale contacts

Renato Buzio*[1], Andrea Gerbi[1], Cristina Bernini[1], Luca Repetto[2], Andrea Silva[3,4] and Andrea Vanossi[3,4]

[1]CNR-SPIN, C.so F.M. Perrone 24, 16152 Genova, Italy
[2]Dipartimento di Fisica, Università degli Studi di Genova, Via Dodecaneso 33, 16146 Genova, Italy
[3]CNR-IOM Consiglio Nazionale delle Ricerche - Istituto Officina dei Materiali, c/o SISSA, Via Bonomea 265, 34136 Trieste, Italy
[4] International School for Advanced Studies (SISSA), Via Bonomea 265, 34136 Trieste, Italy

**ABSTRACT.** Solution-processed few-layers graphene flakes, dispensed to rotating and sliding contacts *via* liquid dispersions, are gaining increasing attention as friction modifiers to achieve low friction and wear at technologically-relevant interfaces. Vanishing friction states, i.e. superlubricity, have been documented for nearly-ideal nanoscale contacts lubricated by individual graphene flakes; there is however no clear understanding if superlubricity might persist for larger and morphologically-disordered contacts, as those typically obtained by graphene wet transfer from a liquid dispersion. In this study we address the friction performance of solution-processed graphene flakes by means of colloidal probe Atomic Force Microscopy. We use an additive-free aqueous dispersion to coat micrometric silica beads, which are then sled under ambient conditions against prototypical material substrates, namely graphite and the transition metal dichalcogenides (TMDs) $MoS_2$ and $WS_2$. High resolution microscopy proves that the random assembly of the wet-transferred flakes over the silica probes results into an inhomogeneous coating, formed by graphene patches that control contact mechanics through tens-of-nanometers tall protrusions. Atomic-scale friction force spectroscopy reveals that dissipation proceeds *via* stick-slip instabilities. Load-controlled transitions from dissipative stick-slip to superlubric continuous sliding may occur for the graphene-graphite homojunctions, whereas single- and multiple-slips dissipative dynamics characterizes the graphene-TMD heterojunctions. Systematic numerical simulations demonstrate that the thermally-activated single-asperity Prandtl-Tomlinson model comprehensively describes friction experiments involving different graphene-coated colloidal probes, material substrates and sliding regimes. Our work establishes experimental procedures and key concepts that enable mesoscale superlubricity by wet-transferred liquid-processed graphene flakes. The proposed methodology is suitable to explore the friction response of other layered nanomaterials prepared by scalable production routes.

## 1. INTRODUCTION

Solution-processed single-layer/few-layers graphene (SLG/FLG) flakes, supplied to macroscale rotating and sliding contacts *via* liquid dispersions, are gaining increasing attention as friction modifiers in practical applications.[1–5] Compared to mechanically cleaved or grown graphene, liquid dispersions appear more suitable for mass production, they do not require time-consuming optimization to coat different material substrates, and they are virtually able to conformally coat interfaces of arbitrary geometry. Whereas the flakes quality strictly depends on the production method,[6] and detrimental effects on their lubricity may arise from unintentional contamination from high-boiling-point solvents and surfactants,[7] solution-processed SLG/FLG flakes probed by AFM were recently shown to display ultralow friction forces comparable with bulk graphite and mechanically cleaved graphene.[7] This points to the promising opportunity to exploit such nanomaterials to achieve ultralow friction states and possibly superlubricity, namely a condition in which the friction force vanishes or very nearly vanishes. It is worth mentioning that the occurrence of almost negligible friction coefficients (i.e. $\ll 0.01$), assessing just the friction variation with the imposed normal load, does not necessarily imply vanishing values of the measured friction force and the overall absence of dissipative stick-slip regimes.[8,9] In general, for graphitic nanosystems, superlubricity develops because of the presence of atomically smooth shear planes at the contact interface. Specifically, nanosized flakes are thought to adhere at the sliding surfaces and arrange in orientationally misaligned configurations, that give almost complete cancellation of the lateral force *via* interfacial incommensurability (structural lubricity).[10–12] Dienwiebel *et al.*[13] first demonstrated experimentally that an individual nanoflake attached to an Atomic Force Microscopy (AFM) tip displays registry-dependent friction when sled against graphite. Later, similar phenomena were reported for other carbon-based nanosystems.[10] There is no clear understanding, however, if superlubricity might persist when nanocontacts are scaled-up in size towards larger but highly disordered contacts,[14] as those typically obtained by depositing SLG/FLG flakes from the liquid phase. Coating micrometric colloidal AFM probes by solution-processed graphene flakes represents an effective strategy to explore this issue. Nowadays, both nanosized (conventional) and colloidal AFM probes have been coated by graphene, either using direct graphene growth,[15–17] triboinduced graphene transfer,[8,9,18–22] all-dry viscoelastic graphene transfer,[23] or graphene wet-transfer from liquid interfaces[24] and from liquid dispersions.[25–28] AFM probes coated by liquid-processed graphene flakes have overall shown reduced interfacial adhesion and friction together with increased lifetime,[28,29] albeit no detailed investigation of their contact interface and of the elementary dissipation mechanisms was attempted. This is likely due to the high degree of uncertainty arising from the broad distribution of thickness and size of the wet-transferred flakes.[30] Such issue contributes to the coating inhomogeneity, together with the variability associated with the random stacking of the transferred flakes over the probes' surface. The actual coating structure thus remains largely uncontrolled in experiments, with a few exceptions.[31–33] In particular Daly *et al.*[33] gained insight on the role of the coating heterogeneous layering, by considering AFM colloidal probes covered by a nanometer-thick (~60nm) multilayer graphene oxide (GO) film. They claimed that the heterogeneous layering of the transferred flakes can critically affect interfacial sliding, as several topological defects do enter the sliding volume and substantially modulate the shear strength. However, as the shear response across GO planes ultimately depends on strong hydrogen-bond networks controlled by intercalated species (hydroxyl/epoxy functional groups and water molecules),[34] the GO-coated colloidal AFM probes were practically unable to explore the emergence of ultralow-friction states. Remarkably, the impact of the jagged morphology of the coating, with nanometer roughness generated by the flakes deposition, was not addressed in conjunction with normal and friction force spectroscopy.

In this study we explore in detail the friction response of mesoscale sliding junctions formed by contacting atomically-smooth model substrates, namely graphite and the TMDs $MoS_2$ and $WS_2$, with graphene-coated colloidal AFM probes. The use of a high-quality, thermodynamically-stable and surfactant-free aqueous dispersion of SLG/FLG flakes allows us to obtain lubricious wet-transferred coatings, without the need to implement high-temperature ($\geq 400°C$) annealing steps to recover the flakes' intrinsic properties.[7] Experiments address how the coating variability impacts contact mechanics and friction. We gain detailed interfacial knowledge through a suitable combination of Scanning Electron Microscopy (SEM) and AFM. Atomic-scale friction force spectroscopy turns out to be excellently described by the single-asperity thermally-activated Prandtl-Tomlinson model, as confirmed by ancillary numerical modelling. The ubiquitous graphitic nanoroughness at the contact interface is shown to effectively link the friction of mesoscopic junctions to a well-established theoretical paradigm of nanotribology.

## 2. Experimental

### 2.1 Characterization of the graphene flakes from a surfactant-free aqueous dispersion

A commercially-available additive-free aqueous dispersion of graphene was used for experiments (named Post-treated 'Eau de Graphène' EdG from Carbon Waters, France). This is a homogeneous stable mixture of single-layer and FLG flakes (typical layers number $\sim 1-8$; broad lateral-size distribution from tens of nanometers up to a few micrometers; pH$\sim 7.5-8.5$) with concentration $\sim 0.1$mg/mL and a shelf life of 3 months at 8°C (graphene precipitation is slowly taking place for longer periods). To obtain the dispersion, potassium graphite $KC_8$ is first exfoliated down to SLG and FLG flakes in tetrahydrophuran THF to yield a thermodynamically stable graphenide (negatively charged graphene) solution. Graphenide ions are then oxidized back to graphene by air exposure, and immediately transferred to degassed water to achieve a remarkably stable solution.[35] In fact, graphene re-aggregation is drastically slowed down in degassed water, by spontaneous adsorption of negatively charged $OH^-$ ions on the hydrophobic graphene surface. To characterize the graphene flakes, we drop-casted ~100μL aliquots of pure EdG solution onto Si(100) wafers terminated with 300nm-thick $SiO_2$ (Crystec GmbH, Germany), used upon ultrasonic bath in acetone and ethanol. After a ~4h sedimentation and drying in ambient conditions, graphene deposits were rinsed with a few drops of DI water (to minimize the KOH residues at the sample surface) and dried with $N_2$ on a 50°C hot plate for a few minutes. Some samples were also prepared by drop casting graphene on Si wafers preliminarily treated by oxygen plasma (30W, 30s, working pressure $p_{O_2} \sim 1.5 \times 10^{-1}$ mbar). Samples were characterized by optical microscopy, AFM and Raman spectroscopy. The morphological AFM imaging was accomplished in contact-mode or tapping-mode (AFM Solver P47-PRO by NT-MDT, Russia equipped with probes CSC37/Al-BS by MikroMasch or OTESPA-R3 by Bruker respectively). Raman spectra were collected with a commercial spectrometer (NRS-4100 by JASCO, Japan), using the 532nm (2.33eV) laser excitation wavelength and × 20 or × 100 objectives. The laser power incident on the samples was ~0.7mW. The spectrometer was calibrated with the G band of HOPG at 1582 $cm^{-1}$. Raman spectra were collected on several spots of the same sample. All peaks were fitted with Lorentzian functions.

## 2.2 Graphene-coated AFM probes: fabrication method and characterization

In order to identify an effective strategy to deposit FLG flakes from the aqueous solution to the AFM probes, we initially considered deposition on both commercial nanoprobes (Si probes HQ:CSC38/Al-BS and HQ:CSC37/Al-BS by MikroMasch; OTESPA-R3 by Bruker; Pt-coated probes HQ:NSC35/Pt by MikroMasch; custom Au-coated probes OMCL-AC160TS by Olympus) and custom-made colloidal probes (with silica beads of diameter ~24μm, prepared as in ref.[8]). Contrary to previous studies,[36,37] we found that the simple immersion of such probes in the graphene solution (from tens of minutes up to several hours) provided erratic results in terms of graphene adsorption, and an overall poor graphene coverage. In particular we were not able in all the performed trials to have FLG flakes covering the very end of the nanoprobes' tip, or the contact region of the colloidal probes (see Supplementary Information Figure S1). Indeed this was the case also when the probes were pre-treated by oxygen plasma.[37] This negative results might reflect crucial differences in the concentration, lateral size or residual functionalization of the EdG flakes compared to flakes synthesized via redox reactions, and dispersed into highly-concentrated (~1 − 10mg/mL) but metastable solutions.[36,37] Therefore, to substantially enhance the graphene coverage using the EdG solution, we implemented a deposition protocol 'mixing' dip-coating and drop-casting techniques. The method is illustrated in Figure 1.

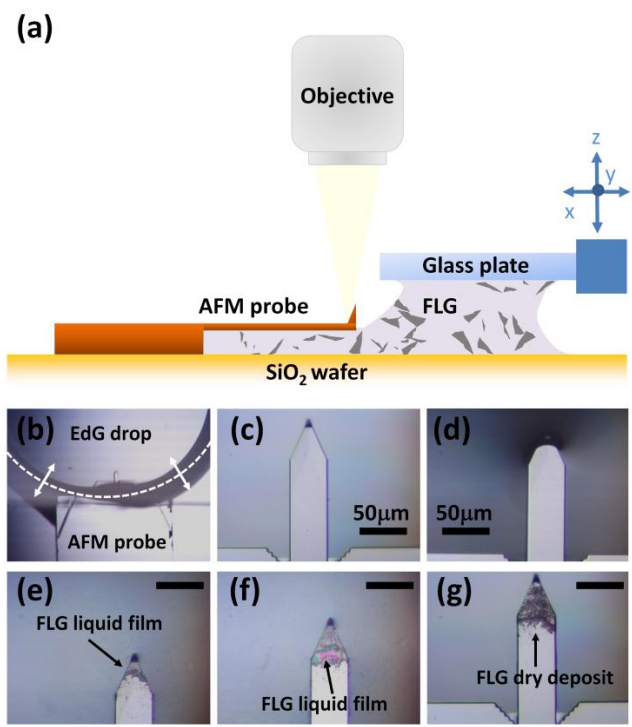

**Figure 1.** Method used to coat AFM probes with graphene from the EdG aqueous solution. (a) Schematics of the experimental set-up (not in scale). (b) Top-view optical micrograph of a commercial nanoprobe, with the cantilever fully dipped into the EdG drop (the dotted line is the position of the drop lateral meniscus driven by the micromanipulator). High-resolution optical micrographs of the same cantilever respectively: (c) pristine uncoated; (d) before retraction of the tipped-end from the EdG drop; (e) coated by a FLG liquid layer, immediately after retraction from the EdG drop; (f) after several 'dip-retract' cycles; (g) coated by a FLG dry deposit.

Briefly, an uncoated AFM probe was firmly placed in contact with a piece of $SiO_2$-terminated Si wafer with the aid of a metal leaf spring. A 10μL drop of EdG solution was dispensed onto $SiO_2$ close to the probe, then a glass plate attached to a XYZ micromanipulation stage (Newport M-460A-XYZ stage with SM-13/DM-13 screws) was slowly lowered until it came in contact with the drop itself. The drop - sandwiched between the glass plate and the Si wafer - was 'pull' by the micromanipulation stage towards the AFM probe, in order to bring the tipped end of the cantilever nearby the drop meniscus (Figure 1a). We note that neither the glass, nor the Si wafer and the AFM probe were pre-treated by oxygen plasma, therefore there was no spreading of the drop over the wetted surfaces (water contact angle ~40° with $SiO_2$). Next, the micromanipulation stage was used to systematically 'dip and retract' the tipped end of the cantilever from the EdG drop (Figure 1b-d). The whole procedure was carried out under an optical microscope, equipped with a long-working distance objective × 50. It is crucial to mention that immediately after retracting the cantilever from the drop, a thin liquid layer usually wetted the cantilever end (Figure 1e-f), so that a new 'dip-retract' cycle was not performed until the liquid layer was fully dried (Figure 1g). This makes the method a 'mixture' of dip-coating and drop-casting techniques. As the whole procedure is prone to exploration and optimization of several parameters, graphene deposition was in practice accomplished as follows. For each probe we carried out one or more 'deposition runs', until a grayish coating appeared on the cantilever tipped end. We identify a 'deposition run' with the total time the 10μL EdG drop remained trapped between the glass plate and Si wafer, before its evaporation. Under standard laboratory conditions (relative humidity $RH = 50 - 60\%$ and temperature $T = 22 \pm 2°C$), the drop evaporated in $\sim 90\ (\pm 30)$ minutes. In the course of this time, one could typically perform ~150 'dip-retract' cycles ($\sim 2 - 3$ cycles/minute) before drop evaporation. At the end of the 'deposition run', the probe was rinsed with DI water, and dried with $N_2$ flow on a 50°C hot plate for a few minutes. A new 'deposition run' started by iterating the procedure above with a new 10μL EdG drop. The rationale was thus to progressively increase the amount of deposited graphene by increasing the total number of 'dip-retract' deposition cycles. The effective coverage on each probe was evaluated *a posteriori* by direct inspection *via* high resolution microscopy. All coated probes were routinely inspected by SEM, completed under a 1kV acceleration voltage using a tungsten filament instrument or a field-emission one (CrossBeam 1540 XB by Zeiss). For colloidal AFM probes, Raman spectra of the graphene coating were collected at $\times 100$ magnification. Additionally, the morphology and friction response of the deposited flakes were studied by reverse AFM imaging on a spiked grating (Tipsnano TGT1), using both dynamic ('tapping') mode and contact mode (normal load $F_N \leq 50\text{nN}$).

### 2.3 AFM measurements with the graphene-coated probes

Normal force and friction force spectroscopies were carried out in contact-mode under ambient conditions. Graphene-coated AFM probes were placed in contact with the freshly-cleaved surfaces of HOPG (grade ZYB by MikroMasch), $2H - WS_2$ or $2H - MoS_2$ crystals (from HQ Graphene) respectively. For the calibration of the elastic constant of each probe $k_C$, and of the normal force $F_N$ and lateral force $F_L$, see our previous publication.[8] Normal force *vs* distance curves ($F_N$ *vs* $D$) were obtained by ramping the probe-sample distance while recording the cantilever deflection signal. We assigned $D = 0$ to the hard-wall repulsion region.[38–40] We obtained atomic-scale $F_f$ *vs* $F_N$ characteristics from friction maps ($512 \times 512$ pixels), in which $F_N$ was systematically decreased every ten lines from a relatively large starting value (i.e. a few hundreds of nanoNewtons depending on the probe) to the pull-off point. To this end, we interrogated surface portions that were free from atomic steps (with a typical scan range $11 \times 11\text{nm}^2$). Friction maps were analyzed in LabView (National Instruments) and they were displayed using the WSXM software.[41]

### 2.4 Atomic-scale friction: modelling and data analysis

We analyzed atomic friction maps by means of the one-dimensional Prandtl-Tomlinson (PT) model[42]. In this framework, a point-like single-asperity – that mimics the AFM tip – is driven over a one-dimensional sinusoidal potential of amplitude $E_0$ and periodicity $a$, representing the corrugated substrate, by means of a pulling spring of value $k$. The spring connects the single-asperity to an external stage moving at speed $v$. The spring indeed represents an effective parameter combining the torsional properties of the cantilever and the mechanical properties of the contact interface. According to the PT model, the single-asperity can move with two distinct regimes that depend on the Tomlinson parameter $\eta = 2\pi^2 E_0/(ka^2)$, i.e. the ratio between the potential amplitude and the effective elastic energy. When $\eta \leq 1$, the total potential energy has a single minimum at any time and the single-asperity moves smoothly over the potential with vanishing dissipation; for $\eta > 1$, two (or more) minima appear in the energy landscape so that the single-asperity dynamics becomes intermittent (stick-slip) and dissipation occurs. Experimentally, we estimated the contact parameters $E_0$, $k$ and $\eta$ as a function of the normal load $F_N$.[8,43] Briefly, for each value of $F_N$, we selected only those specific portions of the experimental lateral force traces having periodicity $a$ (~$0.21 - 0.25$nm for HOPG; ~$0.29 - 0.32$nm for $WS_2$ and $MoS_2$). In fact these correspond to individual slip jumps of the AFM probe approximately along the zigzag crystallographic direction. Smaller (or longer) slips than $a$ were disregarded. For each selected force profile, the corrugation $E_0$, the contact stiffness $k$ and the Tomlinson parameter $\eta$ were estimated as:

$$E_0 = \frac{aF_{L,max}}{\pi} \qquad (1)$$

$$\eta = \frac{2\pi F_{L,max}}{ak_{exp}} - 1 \qquad (2)$$

$$k = \frac{\eta+1}{\eta} k_{exp} \qquad (3)$$

where $F_{L,max}$, $a$ and $k_{exp}$ are the highest local force maxima, the slip distance, and the lateral force slope respectively. To designate the highest force maxima along any selected force trace, we ordered the jumps of slip distance $a$ in terms of decreasing force amplitude $F_L$ and we conventionally assumed $F_{L,max}$ to correspond to the 25% tail of the highest jumps. For each normal load value $F_N$, the mean values of $E_0$, $k$ and $\eta$ were obtained by averaging over an ensemble of several slip jumps (~$100 - 400$). Theoretically, we calculated friction force *vs* displacement profiles by integrating the underdamped Langevin equation:

$$m\ddot{x} + m\gamma\dot{x} = -\frac{\partial V(x,t)}{\partial x} + \xi(t) \qquad (4)$$

where the potential energy $V(x,t)$ reads:

$$V(x,t) \equiv -\frac{E_0}{2}\cos\left(\frac{2\pi x}{a}\right) + \frac{1}{2}k(vt-x)^2 \qquad (5)$$

A 4th-order Runge-Kutta algorithm was used to this end. The instantaneous lateral force trace was evaluated as:

$$F_L = k(vt - x) \qquad (6)$$

For the tip mass we used $m_0 = 1 \times 10^{-12}$ Kg. The Langevin damping $\gamma = 2\sqrt{km} = 0.01$ ns$^{-1}$ was chosen to reproduce experimental force traces. The thermal noise term $\xi(t)$ satisfies the fluctuation-dissipation theorem: $\langle\xi(t)\xi(t')\rangle = 2m\gamma k_B T \delta(t-t')$. A sliding velocity $v = 60$ nm/s and a temperature T = 296K were chosen to mimic experimental conditions. The simulated force traces were computed as averages over tens of stick-slip events in the steady state regime. To reproduce the experimental variation of the tip and sample, simulated forces were obtained as an average over 16 values of $k$ in the range $10 - 40$ N/m and 5 independent realization of the thermal noise. The standard deviation over this average is used as a confidence level in the comparison with experimental data.

## 3. RESULTS and DISCUSSION

### 3.1 Morphology and nanotribology of drop-casted graphene flakes

Drop-casted deposits of FLG flakes on SiO$_2$ showed a discontinuous coverage of graphene patches, as readily appears from the top-view optical micrograph of Figure 2a. Qualitatively, this was the case also when the aqueous dispersion of graphene was drop-cast on SiO$_2$ treated by oxygen plasma (not shown). The inhomogeneity of the deposited material is indeed common to other liquid-phase sources of graphene and it has been reported in previous studies.[7,33] It reflects the relatively low interaction of graphene with the deposition substrate. Magnification of the graphene patches by AFM revealed a micrometric random network, characterized by bare oxide regions alternating with compact agglomerates of flakes. As shown in Figure 2b, the individual flakes mostly stack with their basal plane aligned with the SiO$_2$ surface and are characterized by a broad lateral-size distribution. Friction maps and $F_f$ vs $F_N$ characteristics attested the lubricious response of the flakes, with (ultralow) friction forces by factors $\sim 10 - 20$ smaller than on the uncovered SiO$_2$ and a friction coefficient $\mu \sim 10^{-2}$. It is certainly remarkable that the drop-casted samples show lubricity comparable with bulk graphite upon a short drying step at 50°C in ambient air. In fact, thanks to the additive-free protocol for the preparation of the EdG dispersion, there is no need to implement high thermal annealing to recover the graphene lubricity from contamination by high-boiling-point solvents (e.g. see the 90 minutes annealing at 350 °C in vacuum, to restore lubricity of graphene inks prepared by liquid-phase exfoliation in N-methyl-2-pyrrolidone [7]).

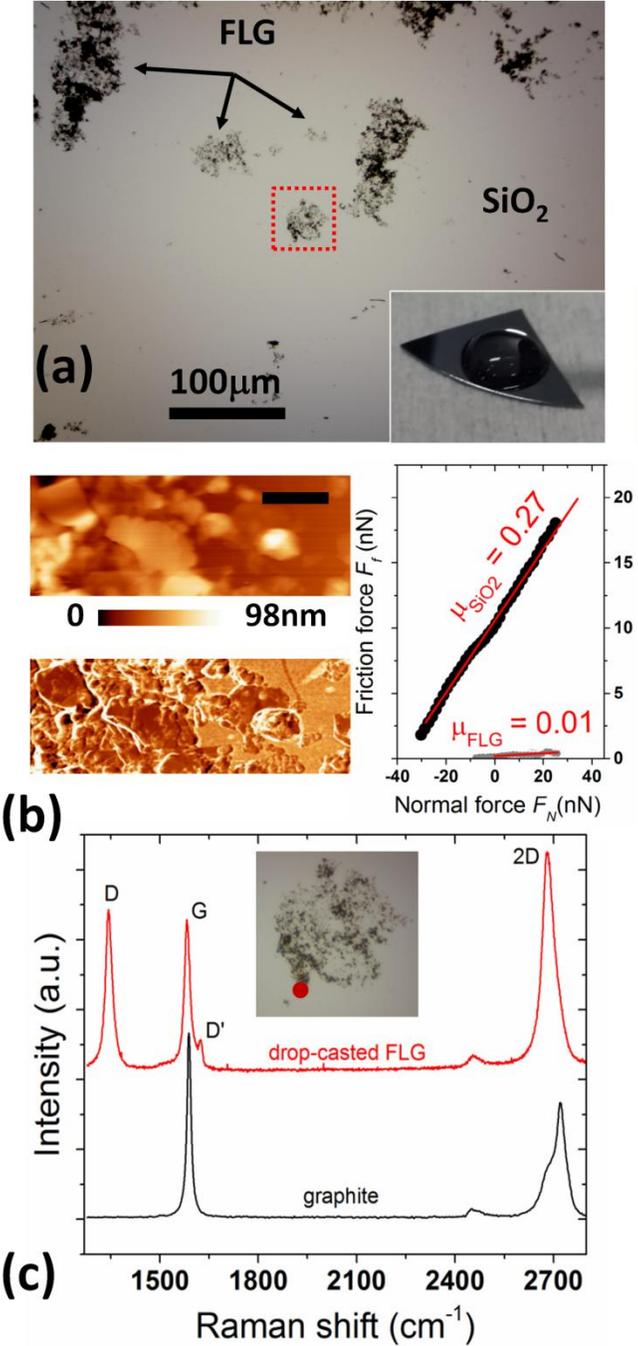

**Figure 2.** (a) Optical micrograph of a graphene patches formed on a $SiO_2$ substrate by drop-casting the water-based FLG flakes dispersion. (b) Topography (top, scale bar 530nm, $F_N = 12$ nN), associated friction map (bottom) and (c) $F_f$ vs $F_N$ curves contrasting the response of FLG

flakes and uncovered $SiO_2$ regions. (d) Representative Raman spectrum acquired on a micrometric region from the FLG patch in (a) (highlighted with the red dotted square).

A representative Raman spectrum for the drop-casted sample is shows in Figure1c. Bands positioned around $1344cm^{-1}$, $1582cm^{-1}$, $1622cm^{-1}$ and $2683cm^{-1}$ correspond, respectively, to the characteristic D, G, D′ and 2D Raman signatures of sub-micrometric few-layer/multilayer graphene flakes. The G peak originates from in-plane vibrational mode involving the $sp^2$ carbon atoms of the graphene sheet ($E_{2g}$ phonon at the Brillouin zone center).[44] Raman bands D and D′ are defect-induced modes observed in disordered graphite and graphene. The D peak is due to the breathing modes of $sp^2$ rings and requires a defect for its activation by double resonance.[45] Double resonance is also at the origin of D′ band. The intensity and width of D and D′ peaks depend on the degree and nature of the basal plane disorder.[44] The 2D band is the second order of the D band. Unlike the D band, however, it does not need to be activated by proximity to a defect, hence it is always a strong band in graphene even when there is no D band present. For an ideal single-layer graphene sheet, both D and D′ bands are absent, whereas the 2D peak is single band ($I(2D)/I(G) \sim 2$). Previous Raman spectroscopy studies on graphene flakes dispersions prepared by liquid phase exfoliation in both aqueous and non-aqueous environments, have shown that besides the G and 2D bands such samples usually show significant D and D′ intensities.[7,46]

For the case of the water dispersion used in the present study, the full-width-at-half-maximum FWHM and the relative intensity of the main bands ($FWHM(D) \approx 23cm^{-1}$, $I(D)/I(G) \sim 1.0 - 1.6$; $FWHM(G) \approx 20cm^{-1}$ ; $FWHM(D') \approx 13cm^{-1}$, $I(D)/I(D') \sim 7.5$; $FWHM(2D) \approx 43cm^{-1}$, $I(2D)/I(G) \sim 1.3 - 1.7$) agree well with those reported by Bepete $et\ al.$[47,48] for graphene thin films stamped on glass and $SiO_2/Si$, after membrane filtration of the EdG dispersion itself. Specifically, both relative intensities $I(D)/I(G) \sim 1.0 - 1.6$ and $I(D)/I(D') \sim 7.5$ can be ascribed to the coexistence of edge defects and basal-plane $sp^3$ point-defects. The latter provide the major contribution, being likely related to some functionalization of the flakes with $-OH$ and $-H$ groups. This is consistent with the evidence that the D/G and D/D′ ratios decrease significantly ($I(D)/I(G) \sim 0.3$, $I(D)/I(D') \sim 3$) when $sp^3$ defects are cured by annealing the thin films at 800°C.[47] Furthermore, as the D peak is relatively narrow ($\approx 23cm^{-1}$) and the D′ peak is not merged with G, the D band certainly reflects to a minor extent the contribution from the edges of the sub-micrometer flakes.[49] Finally, the 2D peak - although broader than in pristine graphene - is still well-fitted by a single Lorentzian.[46] Due to our interest in coating AFM probes by low-temperature processing methods, the $sp^3$ defects were not cured by high-temperature annealing. The amount of defects in nonannealed flakes can be estimated to be in the range $350 - 800$ ppm, that corresponds to a typical distance between point-defects of $\sim 8nm$.[47]

## 3.2 Graphene-coated AFM probes: characterization of morphology and normal-force-spectroscopy

We fabricated graphene-coated AFM probes as described in Figure 1. Representative SEM micrographs of such probes are displayed in Figure 3. It appears that the graphene patches cover different portions of each cantilever, being however mostly concentrated nearby the tipped end due to the controlled micromanipulation of each probe. For sharp probes, evidence of graphene-wrapped nanotips was occasionally found (Figure 3a,b). In such case, normal-force spectroscopy curves measured on HOPG were qualitatively similar to those of the uncoated probes, i.e. they showed sharp snap-in-contact/snap-off-contact and an adhesion force in the range $|F_A| \sim 10 - 20$ nN (see Supplementary Information Figure S2). This agrees with previous reports on graphene-coated nanotips.[16,36]

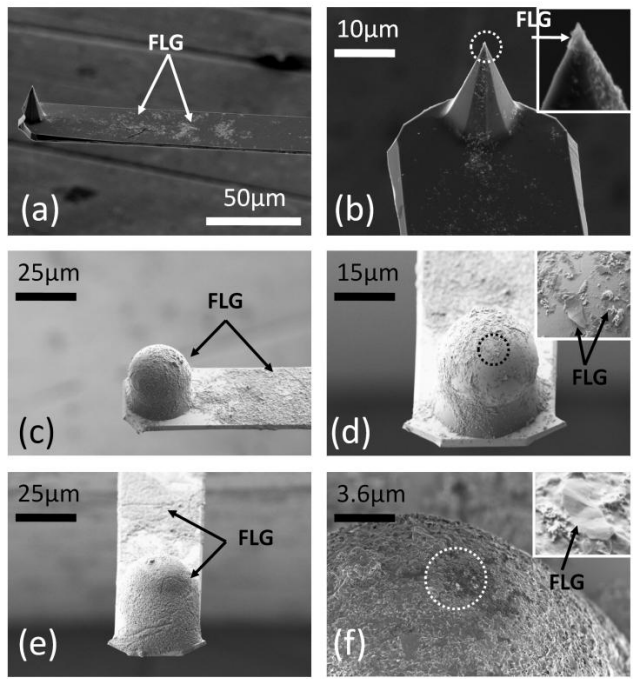

**Figure 3.** SEM micrographs of the graphene-coated AFM probes at different magnifications. (a),(b) Commercial rectangular-shaped silicon cantilever (HQ:CSC37AlBs by Mikromasch), with evidence of the graphene-wrapped nanotip (inset in (b)). (c),(d) Colloidal AFM probe with a silica bead glued onto a rectangular-shaped cantilever, after a total number of 200 'dip-retract' deposition cycles. At higher magnification, the silica surface appears partially covered by FLG flakes (inset in (d)). (e),(f) Colloidal AFM probe as in (c), but after 600 'dip-retract' deposition cycles: the silica surface is coated by a thicker, still inhomogeneous, deposit of flakes. Crumpled flakes are easily discerned at a higher magnification (e.g. see inset in (f), field of view $1.1 \times 1.0$ μm$^2$).

A more complex phenomenology characterized the graphene-coated colloidal probes. High-resolution SEM micrographs showed that the graphene coating was far from homogeneous at the sub-micrometric length scale, with uncoated silica regions alternating with tens-of-nanometers tall protrusions formed by randomly stacked and/or highly-crumpled flakes (Figure 3c-f). This morphology is corroborated by reverse AFM topographies (see Supplementary Information Figure S3) and agrees with that discussed for the drop-casted samples (Figure 2a,b). Likewise, Raman spectra collected on the graphene-coated beads were comparable with those of the drop-casted specimens (see Supplementary Information Figure S4). We point out that the observation of graphene patches over the beads' surface does not ensure, by itself, the manifestation of graphene-mediated effects in contact mechanics. In fact, for the micrometric beads used in the present study (nominal diameter $\sim 24\mu m$), the circular contact spot with an ideally-smooth countersurface has a diameter of about $2a_0 \sim 300nm$ (see Supplementary Figure S1 in[9] for an estimate of the contact radius $a_0$ using a sphere-on-flat contact mechanics theory). Hence, graphene-mediated contact phenomena may appear provided that the graphene patches either exactly coat such contact spot or – which is more often the case – they prevent the silica-substrate contact by forming a new, off-centered and topographically highest contact asperity. We exemplify this issue by considering the results summarized in Figure 4.

Here, the surface morphology of an ideally-smooth silica bead was systematically characterized by AFM in the course of four successive 'deposition runs' (Figure 4a-j), together with the evolution of the adhesive force $F_A$ against HOPG (Figure 4k). Note that each 'deposition run' here consists of only 50 'dip-retract' deposition cycles. One can see that upon runs 1 to 3, the deposited FLG flakes neither coat the mesoscopic contact spot nor they are thick enough to generate a new contact asperity (Figure 4c-h). Thus the adhesion between the colloidal bead and HOPG is controlled by direct $SiO_2 - HOPG$ contact and turns out to be comparable to the pristine case, i.e. $F_A \sim 2500 - 3000nN$. However, an adhesion breakdown to $\sim 500nN$ takes place after 'deposition run' 4, as in this case the newly deposited FLG flakes do contribute to form the topographically highest contact asperity (Figure 4i,j). The agglomerated flakes have thickness of several tens of nm and maintain a lubricious behavior (Figure 4l). More importantly, they have an irregular morphology that reflects their random pile-up and non-conformal adhesion to the silica surface (see also Supplementary Information Figure S5). Accordingly, the key role of the deposited FLG flakes is to generate a lubricious nanoroughness over the surface of the colloidal beads, that drives both the decrease of the contact area (from meso to nanoscale) and of the contact forces. Analysis of representative AFM topographies with threshold criteria gives a rough estimate of $\sim 2 \times 10^2 nm^2$ for the contact area at the topographically highest contact spot with graphite (Supplementary Information Figure S6). This phenomenology is well documented in colloidal probe AFM experiments involving nominally rigid interfaces.[50–52] Even more remarkable, adhesion breakdown was also shown to occur when the nanoroughness originates from tribo-induced material transfer of FLG flakes from a graphitic substrate to the sliding colloidal probe.[8]

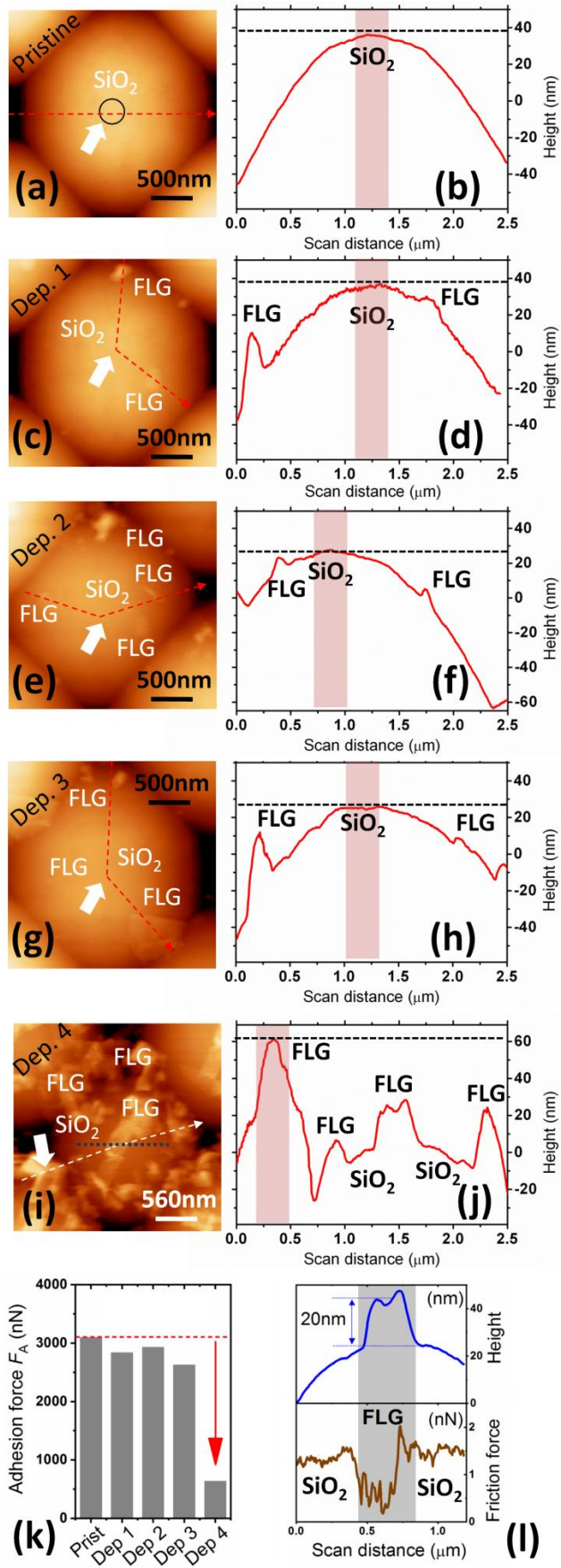

**Figure 4.** (a) AFM morphology and (b) cross-section height (along the dash line in (a)) for a pristine silica bead. The white arrow and dotted circle in (a) highlight respectively the position and size of the contact spot with HOPG (see text). (c)-(j) Evolution of the surface morphology and of the cross-sectional height for the same bead upon four 'deposition runs'. Only after 'deposition run' 4, the topographically highest contact asperity becomes off-centered and located over a graphene deposit (see (i),(j)). (k) Adhesion force $F_A$ measured on HOPG after each 'deposition run': adhesion breakdown occurs after the fourth run. (l) Cross-section height and friction force along the dotted line in (i): it shows the lubricious behavior of deposited FLG flakes compared to SiO$_2$.

Figure 5 further clarifies the impact of the graphene coating on $F_N$ vs $D$ curves acquired on HOPG.

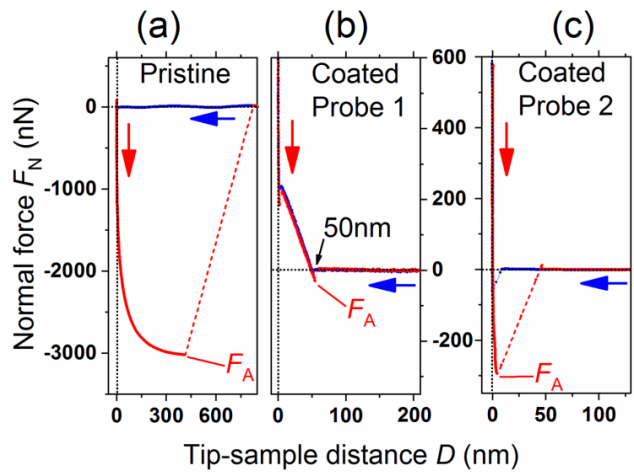

**Figure 5.** (a) Normal force *vs* distance curve on HOPG for a pristine colloidal probe. (b) Normal force *vs* distance curve for a graphene-coated colloidal probe named 'coated probe 1'. Besides adhesion reduction, a long-ranged repulsive interaction at ~50nm signals the response of the elastically-soft graphene coating. (c) As in (b) but for a different probe named 'coated probe 2'. Adhesion is still reduced compared to the pristine contact in (a), but there is no evidence of the coating compliance.

Figure 5a,b contrasts the typical response of a pristine silica bead with that of a graphene-coated bead (named 'coated probe 1') and corresponding to the probe of Figure 3e,f. As previously mentioned, the most prominent effect of the graphene coating is to induce the adhesion breakdown, here from $F_A$~3000nN to ~30nN. Besides this, a fully-reversible long-ranged repulsive interaction arises at a few tens of nanometers distance from the hard-wall-repulsion contact line $D = 0$. We ascribe this peculiar feature to an elastic contribution, due to some mechanical compliance of the (rough) graphene coating in the low-loads regime. Several concurrent factors might contribute to this response. In fact, the flakes do not necessarily stack to form a compact overlayer and load-induced deformations of the graphene coating might take place at low loads. A similar effect was previously observed for graphene-coated sharp AFM probes prepared by dip coating,[32] this being ascribed to the existence of nanogaps between the flakes and the coated surface. Also, $F_N$ vs $D$ curves similar to ours were reported by Ishikawa et al.[25] for individual micrometric flakes partially attached the colloidal beads. The long-ranged repulsion was ultimately explained by a progressive flattening of the attached flakes on increasing the normal load. Consistently, such compliant response was not observed in $F_N$ vs $D$ with graphene flakes tightly attached to the probe surface, i.e. directly glued to the colloidal bead.[23] As such instabilities in normal force and friction force do depend on the random stacking features of the graphene coating, we observed qualitative variations from probe to probe. Figure 5c shows the $F_N$ vs $D$ spectroscopy curve for a different probe (named 'coated probe 2'), which in fact reveals adhesion reduction compared to the pristine contact, albeit there is no evidence of any elastically-soft response of the coating in this case. Consistently with the previous picture, variations in the long-ranged behavior of the spectroscopic curves were occasionally observed in the course of experiments conducted with the same probe, this being a behavior consistent with the release of loosely attached FLG flakes from the coated probe to the contact substrate (Supplementary Information Figure S7). We show below that despite unavoidable differences among $F_N$ vs $D$ curves of the coated probes, elementary dissipation mechanisms on atomically smooth substrates could be effectively rationalized within the framework of the single-asperity thermally-activated PT model.

### 3.3 Atomic-scale friction and superlubricity of graphene-coated AFM probes

The elementary dissipation mechanisms were explored through load-dependent atomic-scale friction force spectroscopy. Representative lateral force maps acquired respectively on HOPG, $WS_2$ and $MoS_2$ by means of the 'coated probe 1' (see section 3.2) are shown in Figure 6a-c. They attest that the sliding motion was typically of stick-slip type. Indeed this occurred over a broad range of normal loads ($0 \lesssim F_N \lesssim 700$nN); hence dissipation at the FLG/HOPG, FLG/$WS_2$ and FLG/$MoS_2$ interfaces was controlled by an atomic-scale interlocking mechanism, taking place between the topographically highest asperity of the graphene coating and each atomically-smooth substrate. According to the $F_f$ vs $F_N$ characteristics in Figure 6d, the smallest friction occurred at the FLG/HOPG interface, intermediate friction at the FLG/$WS_2$ interface and the highest dissipation at the FLG/$MoS_2$ one. This was the case for normal load values both below or above the jumps signaling the peculiar transition of the 'coated probe 1' in normal-force spectroscopy (see Figure 5b). The trend $F_f(\text{FLG/HOPG}) < F_f(\text{FLG/}WS_2) < F_f(\text{FLG/}MoS_2)$ qualitatively agrees with the outcome of several ambient AFM experiments,[19,26,53–55] conducted with nanosized probes on single-layer, few-layer and bulk substrates of graphite, $WS_2$ and $MoS_2$ respectively. Albeit a comprehensive understanding of such trend is missing, it is likely that the different out-of-plane elasticity of the three substrates ($E_{\text{HOPG}} \approx 38$GPa, $E_{\text{MoS2}} \approx 52$GPa, $E_{\text{WS2}} \approx 60$GPa)[54] and the extreme friction sensitivity to ambient humidity of TMDs[56,57] compared to graphite,[58] do dictate the splitting of the friction forces.

The weak dependence of $F_f$ on $F_N$ results in ultrasmall friction coefficients ($\mu_{TL/WS2} \sim \mu_{TL/Mo2} \lesssim 5 \times 10^{-4}$) for WS$_2$ and MoS$_2$, and a slightly higher value ($\mu_{TL/HOPG} \sim 3 \times 10^{-3}$) for HOPG. However, a condition of nearly vanishing friction strictly occurred only for the FLG/HOPG homojunction under the lowest (tensile) loads, i.e. $F_N = -21$nN. This was in fact clearly signaled by the evolution of the lateral force loops, from (dissipative) stick-slip to continuous (superlubric) sliding, on reducing the load $F_N$ (Figure 6e). This load-dependent phenomenology is well documented in single-asperity AFM studies on graphite, TMDs,[59,60] and other atomic and molecular crystals,[42,61] and suggests to rationalize the response of the graphene-mediated contact on the three substrates within the single-asperity PT model. The load-dependent variation of the interfacial parameters $E_0$, $k$ and $\eta$ extracted from the $F_f$ vs $F_N$ curves of Figure 6d, is resumed in Figure 6f-h. Specifically Figure 6f reveals that the potential corrugation $E_0$ depends weakly on $F_N$ and assumes the highest values for the FLG/MoS$_2$ and FLG/WS$_2$ interfaces, being at least a factor ~2 smaller for the FLG/HOPG case. For the latter, $E_0$ varies from ~3eV ($F_N$~600nN) to less than ~1eV ($F_N < 100$nN). Figure 6g shows that the lateral contact stiffness $k$ varies less from one interface to the other, being in the range $30 - 45$N/m for $F_N > 100$nN. As a result, the Tomlinson parameter $\eta \propto E_0/k$ acquires different values for the FLG/HOPG, FLG/MoS$_2$ and FLG/WS$_2$ interfaces, that mostly reflect the splitting of the interfacial potential corrugation $E_0$ among the three systems (Figure 6h). Through renormalization of the friction force by $ak$, $F_f$ vs $F_N$ curves can be mapped into the adimensional $F_f^*$ vs $\eta$ curves, with $F_f^* \equiv F_f/ak$ (Figure 6i). This graph demonstrates that the friction response of the graphene-coated probe follows very well the predictions from the thermally-activated PT model at T = 296K. Importantly the graph includes atomic-scale friction data acquired with multiple graphene-coated probes, which strengthens the generality of our results (see also Supplementary Figures S6, S7). The friction response of the layered contact junctions, embodied by the $F_f$ vs $F_N$ characteristics of Figure 6d, is thus reconducted to contact pinning effects. In fact, pinning is enhanced when going from FLG/HOPG to FLG/WS$_2$ and FLG/MoS$_2$. As in the latter two the effective corrugation potential $E_0$ (and thus the Tomlinson parameter $\eta \propto E_0$) is higher, the layered heterojunctions fall into the highly-dissipative stick-slip regions of the $F_f^*$ vs $\eta$ plot, although characterized by very small values of the coefficient of friction. For the FLG/HOPG homojunction, however, nearly continuous superlubric sliding is possible since, for such interface, $\eta$ assumes the smaller values ($2.0 \lesssim \eta \lesssim 5$) and the contact may transition from stick-slip to continuous superlubric sliding thanks to a load-controlled reduction of $\eta$. Note that the working temperature of the experiments T = 296K allows the transition to be observed around $\eta \sim 2$ (thermolubricity), rather than at $\eta = 1$ as predicted by the athermal (T = 0K) PT model. On the other hand, at $\eta \gtrsim 10$ the contact heterojunctions enter a multi-slip regime.[59] This change is indicated by the clear downward flexion of the theoretical curve in Figure 6i, which follows well the experimental data. In this underdamped, multi-slip regime the details of the contact become progressively more important as $\eta$ increases, resulting in a sizable spread of the experimental data and increased confidence level of the theoretical curve. Moreover, the origin of the double slips is rooted in the underdamped nature of the contact. In this regime each slip is accompanied by a larger force drop due to the large change in the spring elongation. Thus different values of $k$ result in different recorded forces.

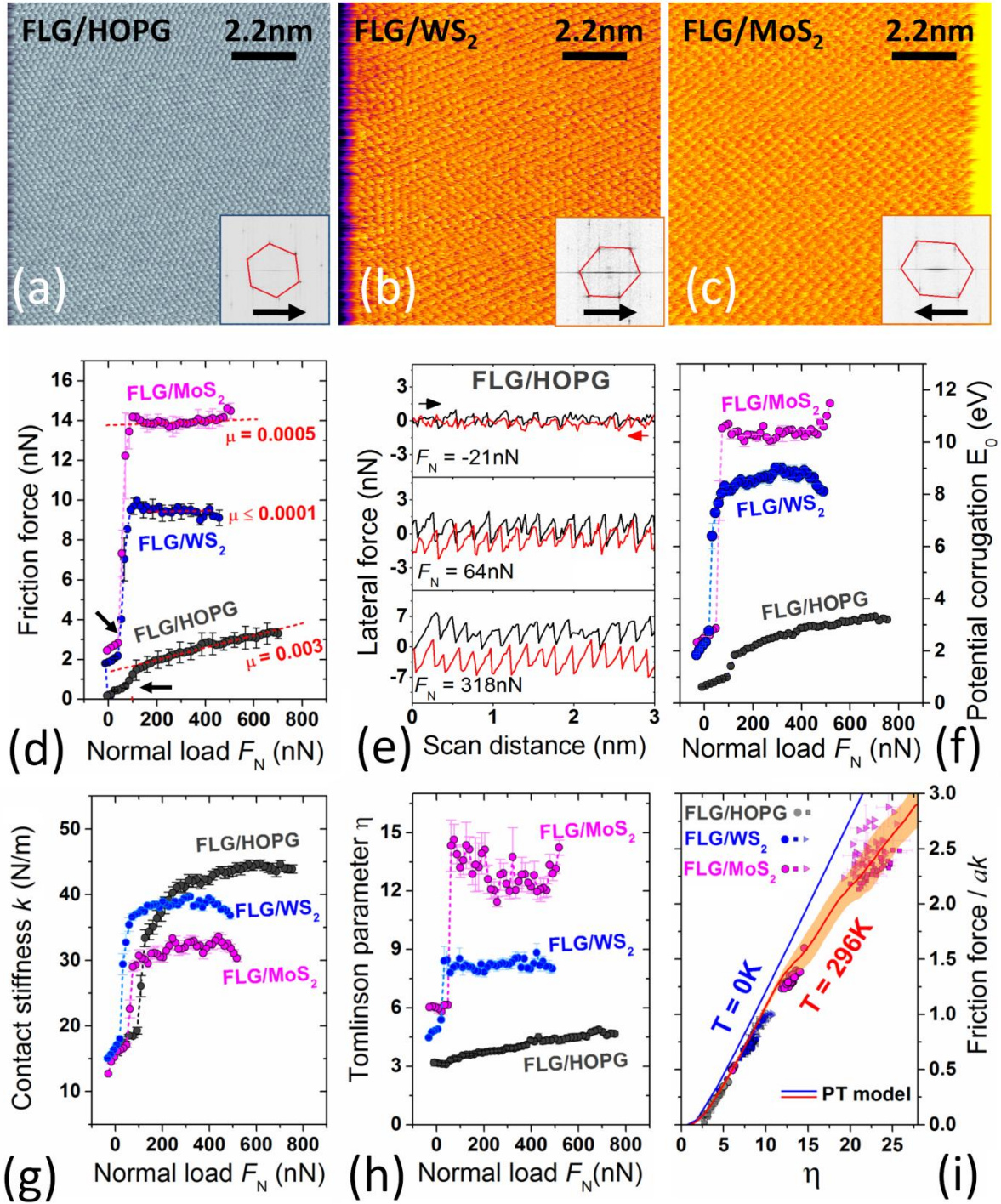

**Figure 6.** (a)-(c) Atomic-scale lateral force maps acquired respectively on HOPG ($F_N = 370$nN), WS$_2$ ($F_N = 205$nN) and MoS$_2$ ($F_N = 218$nN), using the 'coated probe 1' of Figure 5. (d) Set of representative $F_f$ vs $F_N$ characteristics measured for the three layered sliding junctions ($v = 33$nm/s). The sharp friction jumps (highlighted by black arrows) reflect the specific normal-force response of the used probe. (e) Load-

dependent friction loops for the FLG/HOPG interface in (d). (f)-(h) Load-dependent variation of the interfacial parameters $E_0$, $k$ and $\eta$, extracted from the friction characteristics in (d). (i) Comparison of experimental $F_f^*$ vs $F_N$ data with predictions from the PT model, for three different graphene-coated colloidal probes (indicated respectively by circles, squares and triangles). The PT confidence level is shown in orange.

Hence, despite reasonable differences from probe to probe originated by the variability of the graphene coating morphology and compliance, the single-asperity PT model provides a fruitful and comprehensive framework to describe graphene-mediated sliding friction.

The phenomenology depicted above shares several similarities with the case of colloidal AFM probes coated by triboinduced graphitic transfer layers.[8,9] There, atomic-scale friction between the tribo-induced FLG flakes and graphite is governed by an individual nanocontact that corresponds to the highest triboinduced nanoasperity. As a result, (i) atomic friction depends on the energy landscape experienced by such nanoasperity and load-controlled transitions from dissipative stick-slip to continuous superlubric sliding are possible according to the PT model.[8] Additionally, (ii) dissipation systematically increases when the graphite substrate is replaced with $WS_2$ or $MoS_2$.[9] Furthermore, (iii) as soon as the flakes are tribotransferred to the bead surface, one observes a breakdown of interfacial adhesion compared to the smooth-bead case. The present study clearly extends previous findings[8,9] to the situation in which a state-of-the-art dispersion of graphene flakes is wet-transferred at the contact interface. Quite interestingly, the observation of the superlubric transition for the FLG/HOPG homojunction underlines the mandatory use of an additive-free dispersion, but it also gives direct proof that the small amount of topological defects of the solution-processed flakes (basal-plane $sp^3$ point-defects with $-OH$ and $-H$ functionalization and concentration $\leq 800$ppm, see section 3.1) does not enhance contact pinning effects compared to the triboinduced-flakes case. This is supported by the evidence that at low loads $F_N < 100$nN we estimate an interfacial potential corrugation $E_0 < 1$eV for both tribotransferred[8,9] and solution-processed flakes. The tight correspondence between the two systems thus points to the ubiquitous role of interfacial graphitic nanoasperities as key players for solid lubrication in the explored friction regime. The widespread applicability of our results is related to the fact that nanoroughness not only characterizes the wet-transferred graphene flakes prepared in this work, but also the morphology of 2D flakes transferred on solid supports by means of other protocols. In fact, wet-transferred flakes commonly display heterogeneous stacking, as well as folds, cracks and wrinkles.[62] Hence, our study elucidates key concepts of relevance for different contact junctions - lubricated either by liquid dispersions of graphene or by other wet-transferred 2D flakes[63–65] - that are operated under nominally wearless sliding conditions or do involve more complex tribochemistry and third-body-effects.[5,66]

## 4. CONCLUSIONS

In summary, we addressed the nature of the elementary energy dissipation mechanisms, together with the appearance of superlubricity, in sliding contacts lubricated by wet-transferred graphene flakes. Our tribological system comprises graphene-coated colloidal AFM probes sled against graphite, $MoS_2$ or $WS_2$ single-crystal substrates under ambient conditions. We show that the random stacking of the wet-transferred FLG flakes gives sizeable differences in the load-bearing capacity from probe to probe, albeit a prominent reduction of interfacial adhesion is always observed for all the coated probes. This naturally arises from the coating nanoroughness, that reduces the effective contact area between the coated probe and the substrates from meso- to nano-scale. We also demonstrate that energy dissipation occurs *via* atomic-scale stick-slip instabilities, likely governed by the potential energy landscape experienced by one dominant graphitic nanocontact. Nearly continuous superlubric sliding is observed under low loads for graphitic homojunctions, indicating that in such case interfacial pinning phenomena may disappear in favor of structural lubricity effects. This description shares relevant similarities with the superlubricity machanism recently elucidated for graphene-coated probes prepared by triboinduced material transfer, thus contributing to define a general framework for solid lubrication by FLG flakes. The thermally-activated single-asperity PT model is shown to provide a comprehensive description of the main experimental results. Our work establishes experimental procedures and key concepts that enable superlubricity by wet-transferred liquid-processed graphene flakes. The proposed methodology is suitable to address the friction performance of other nanomaterials prepared by scalable production routes targeting hi-tech mechanical applications.

## ASSOCIATED CONTENT

**Supporting Information**

The Supporting Information is available free of charge at https://pubs.acs.org/doi/XXXXXXXX

> SEM micrographs of AFM probes prepared by dip coating (Figure S1), normal force spectroscopy of a graphene-wrapped AFM nanoprobe (Figure S2), AFM morphologies of graphene-coated colloidal AFM probes (Figure S3), optical micrographs and Raman spectra of a graphene-coated colloidal probe (Figure S4), AFM and SEM micrographs of the contact region for a graphene-coated colloidal probe (Figure S5), contact area for the topographically highest nanoasperity (Figure S6), variation of the normal force spectroscopy upon release of a graphene flake from a colloidal probe (Figure S7), atomic-scale friction force spectroscopy for two different graphene-coated colloidal probes (Figures S8 and S9).

**AUTHOR INFORMATION**

*Corresponding Author

Renato Buzio - Email: renato.buzio@spin.cnr.it

**ACKNOWLEDGEMENTS**

R.B. acknowledges useful discussions with Thomas Bottein, Charlotte Gallois and Elodie Jobert from Carbon Waters, and with Carlos Drummond from CNRS CRPP and Université de Bordeaux. This work was financially supported by the MIUR PRIN2017 project 20178PZCB5 "UTFROM - Understanding and tuning friction through nanostructure manipulation". A.V. acknowledges also support from ERC Advanced Grant ULTRADISS, contract No. 86344023.

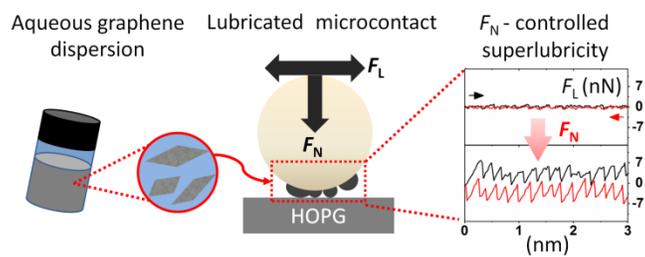

**Table of Contents TOC**

# Solid lubrication by wet-transferred solution-processed graphene flakes: dissipation mechanisms and superlubricity in mesoscale contacts

Renato Buzio*[1], Andrea Gerbi[1], Cristina Bernini[1], Luca Repetto[2], Andrea Silva[3,4] and Andrea Vanossi[3,4]

[1] CNR-SPIN, C.so F.M. Perrone 24, 16152 Genova, Italy
[2] Dipartimento di Fisica, Università degli Studi di Genova, Via Dodecaneso 33, 16146 Genova, Italy
[3] CNR-IOM Consiglio Nazionale delle Ricerche - Istituto Officina dei Materiali, c/o SISSA, Via Bonomea 265, 34136 Trieste, Italy
[4] International School for Advanced Studies (SISSA), Via Bonomea 265, 34136 Trieste, Italy

# Supplementary Information

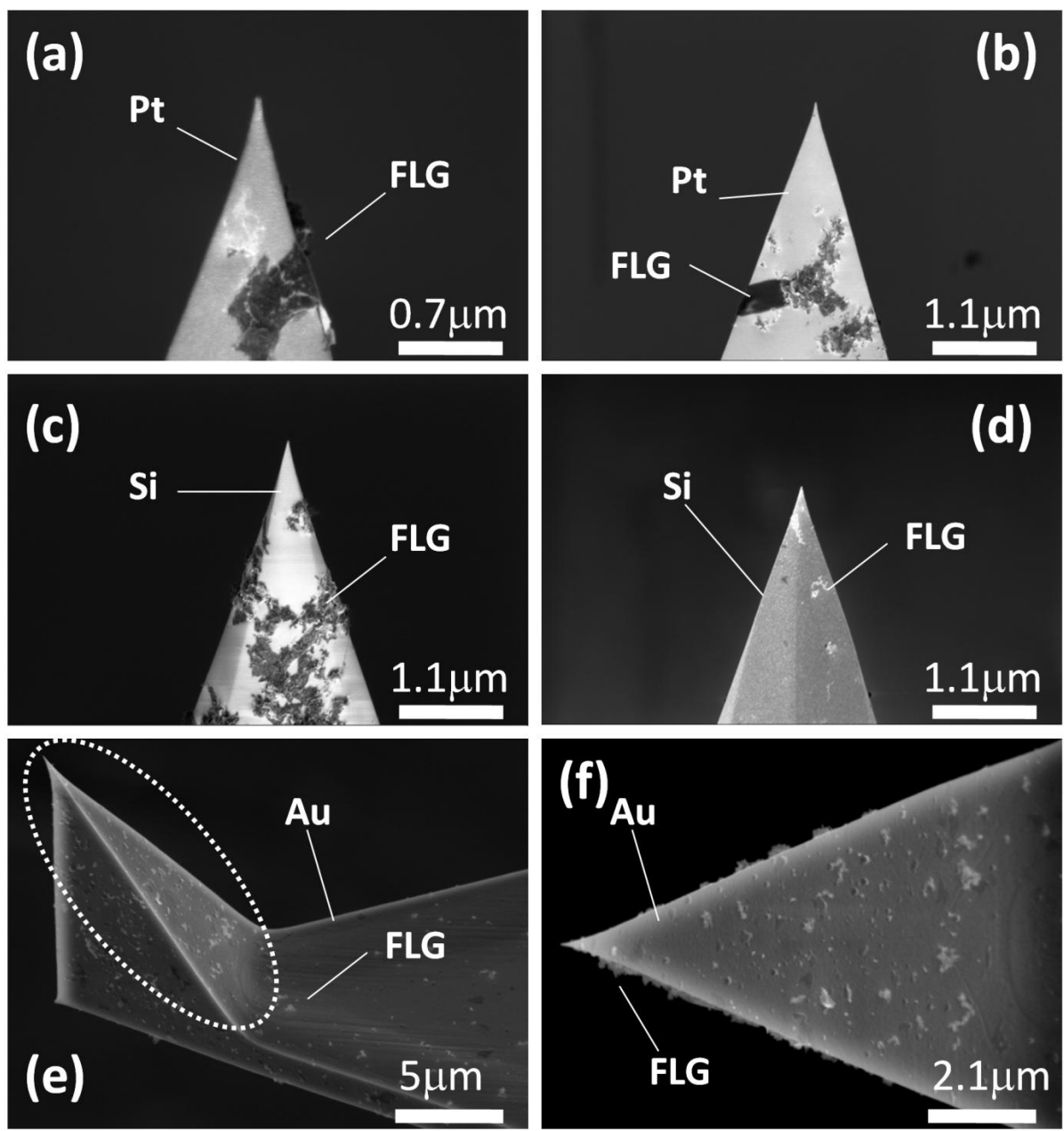

**Figure S1.** SEM micrographs of commercial AFM probes dipped into Eau de Graphene solution. Despite variations of the preparation protocol and probes material there was no evidence for graphene-wrapped tips, albeit FLG flakes were clearly discerned over different regions of the probes' surface. (a),(b) Pt-coated probes (HQ:NSC35/Pt by MikroMasch), treated by oxygen plasma (30W, 30s, working pressure $p_{O_2} \sim 1.5 \times 10^{-1}$ mbar) prior to immersion into graphene solution (dipping time 36 minutes). (c),(d) Si probe (HQ:CSC38/Al-BS by MikroMasch), treated by oxygen plasma (as in (a)-(b)) and dipped into graphene solution (dipping time 100 minutes). (e) Au-coated probe (custom OMCL-AC160TS by Olympus) dipped into graphene solution (dipping time 40 minutes). (f) Magnification of the dotted region in (e).

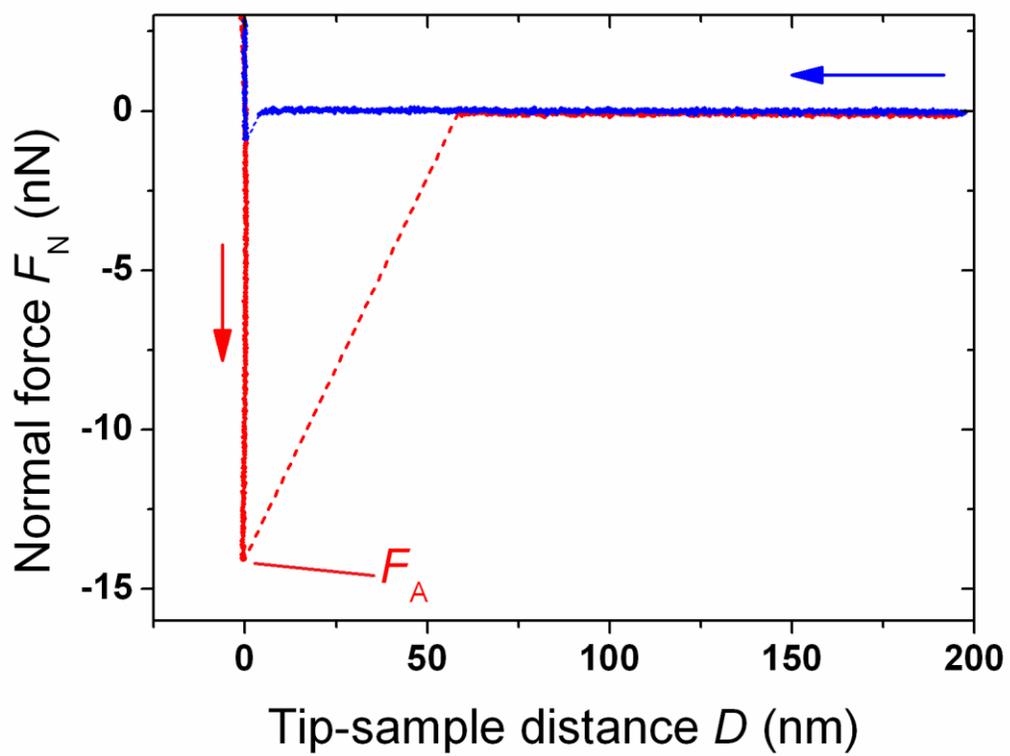

**Figure S2.** Normal force *vs* distance ($F_N$ *vs* $D$) curve acquired on HOPG with the graphene-wrapped AFM probe of Fig. 3(a),(b) (see main text). The curve shows sharp snap-in-contact and snap-off-contact, and an adhesion force $F_A \cong -14$nN.

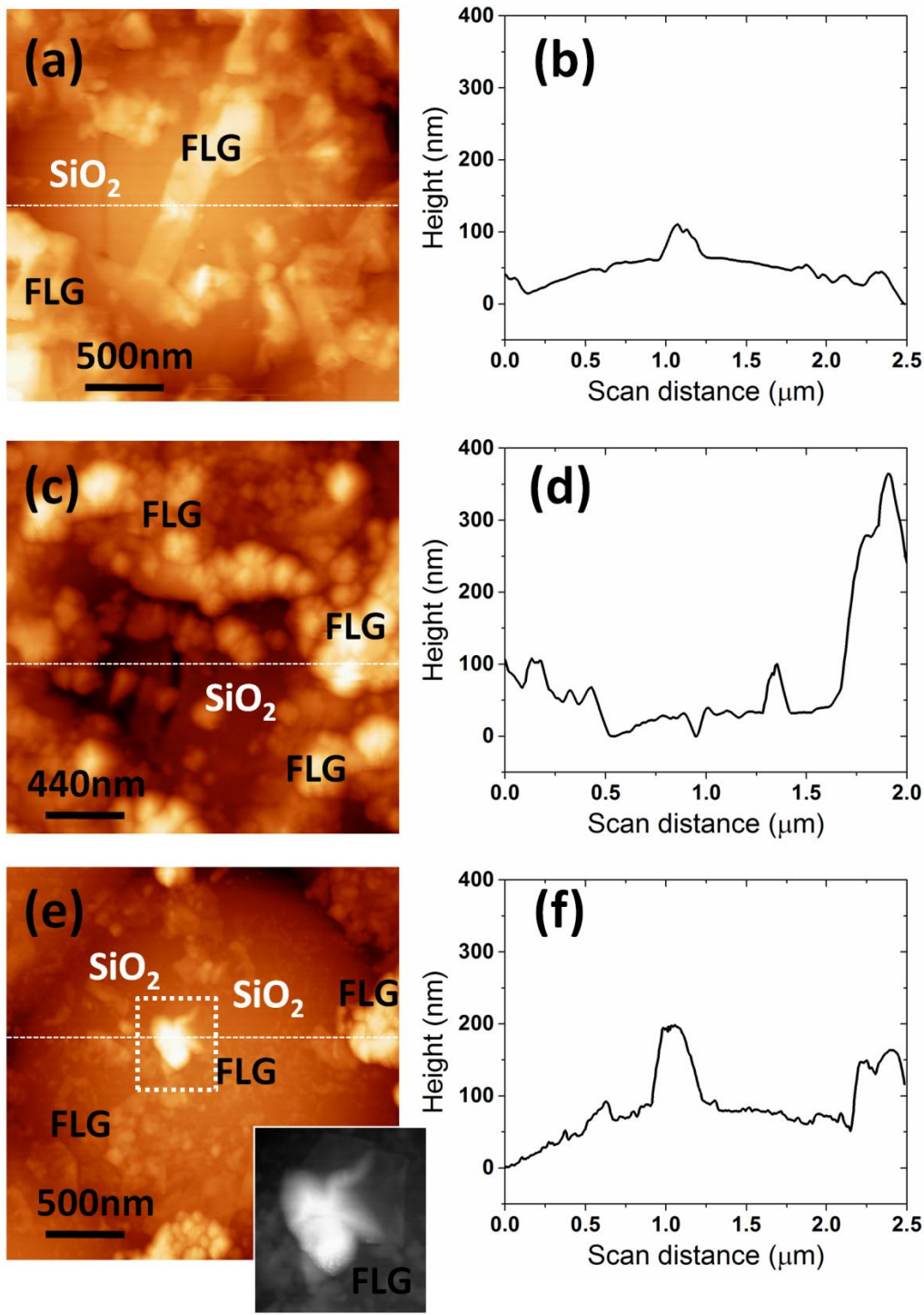

**Figure S3.** Reverse AFM morphologies (left column) and cross-section heights (along the dash lines; right column) for three silica beads coated by wet-transferred flakes. (a-b) They correspond to the colloidal bead of Figure 3c,d and Figure 4 in the main text. (c-d) They correspond to the colloidal bead named "Coated Probe 1" (see Figure 3e,f, Figure 5b and Figure 6 of the main text). (e-f) They correspond to the colloidal bead named "Coated Probe 2" (Figure 5c of the main text).

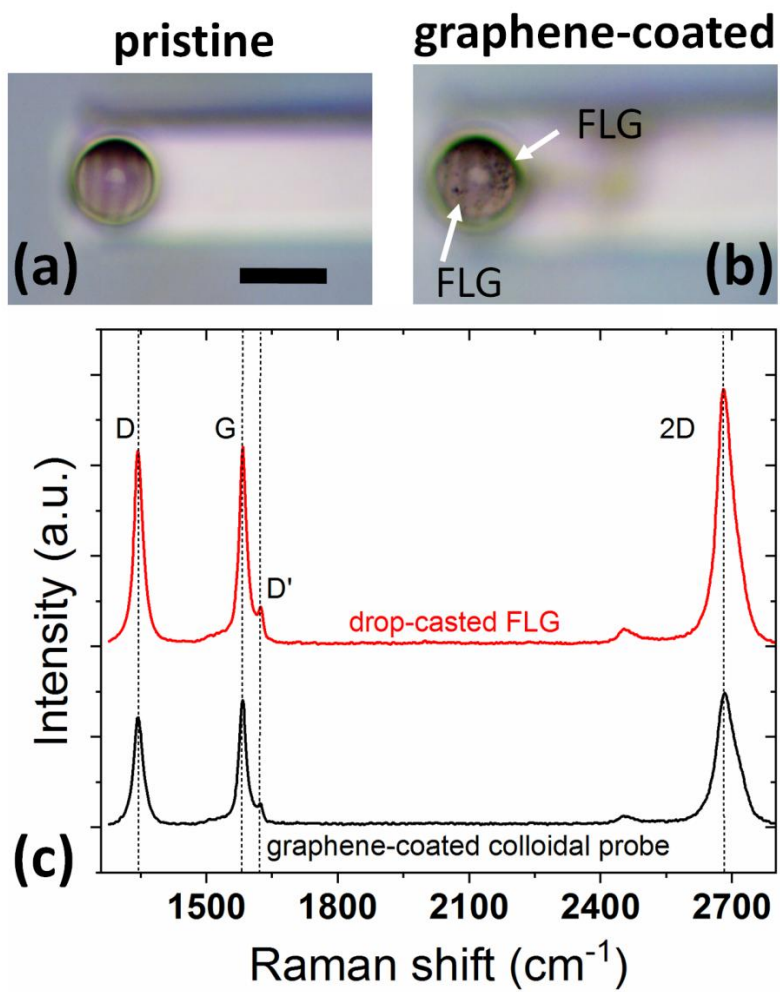

**Figure S4.** Optical micrographs of (a) pristine colloidal AFM probe and (b) the same probe after being coated by FLG flakes (scale bar in (a) is 27μm)). (c) Comparison of Raman spectra collected respectively on drop-casted FLG flakes (red curve) and at the surface of the graphene-coated probe (black curve).

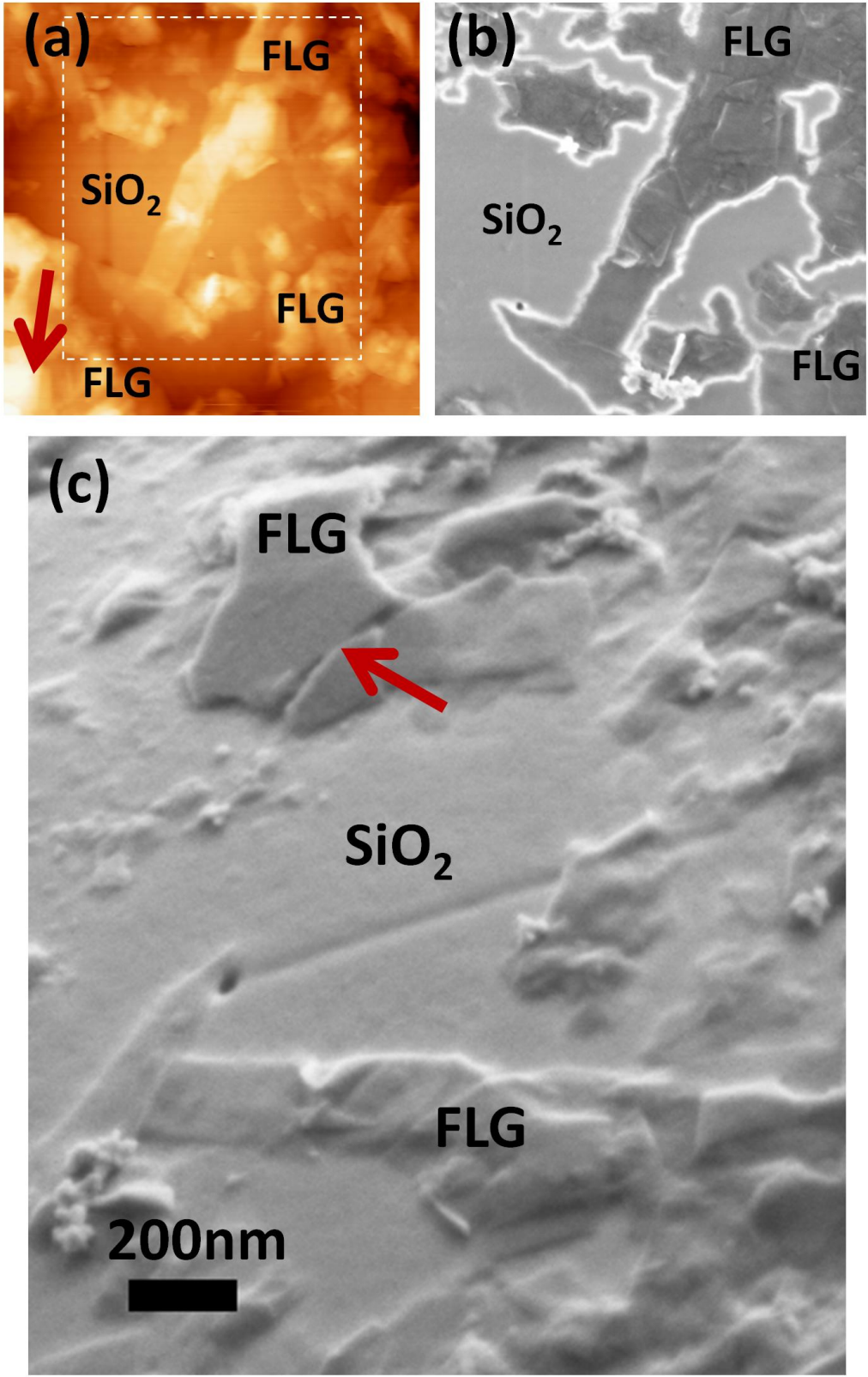

**Figure S5.** (a) AFM morphology acquired by reverse imaging nearby the contact region of the graphene-coated colloidal probe of Figure 3c,d and Figure 4i. (b),(c) SEM micrographs of the same region, from different viewpoints. The arrow highlights the location of the topographically-highest contact asperity.

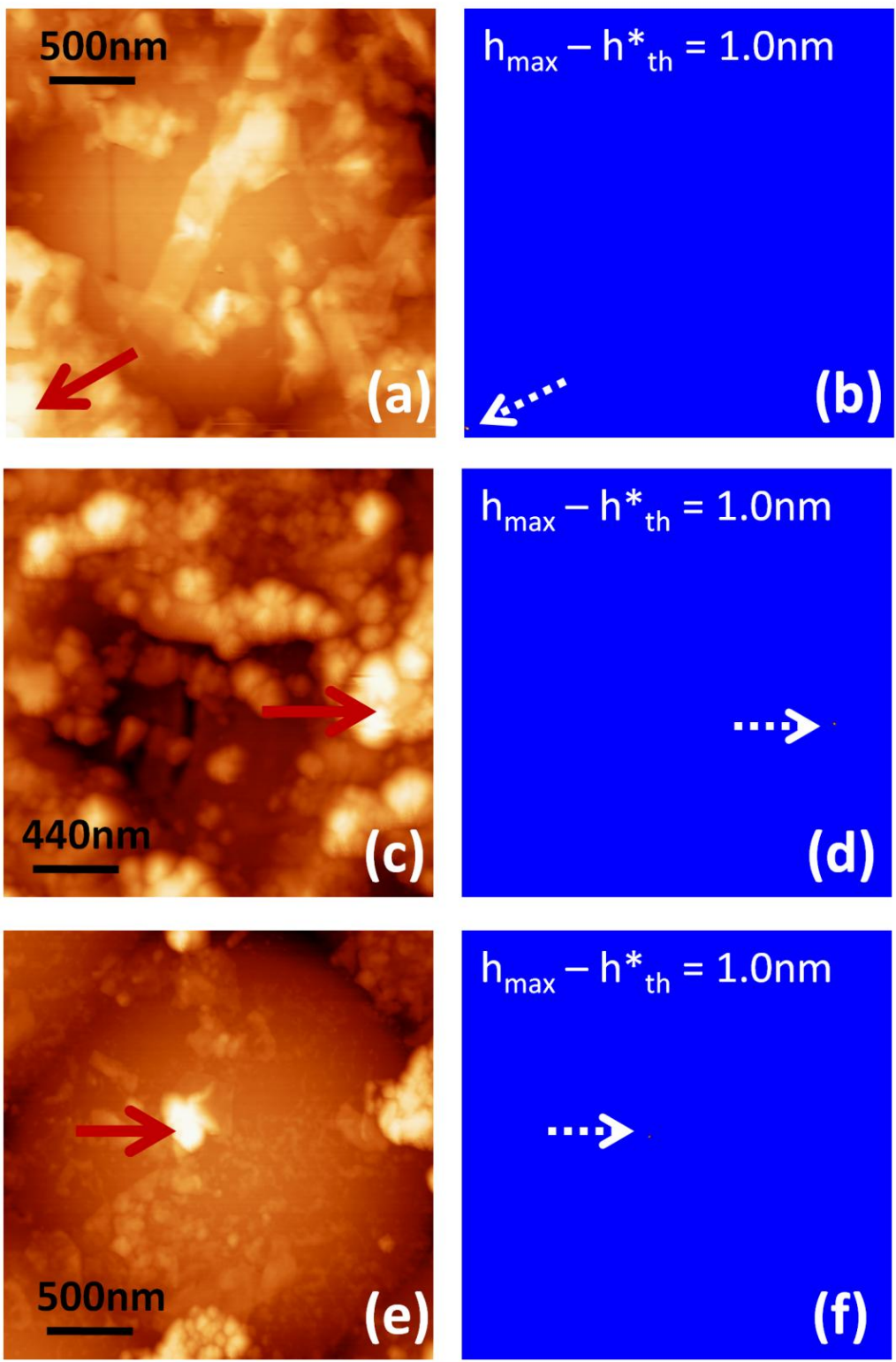

**Figure S6.** Estimation of the area of the topographically-highest nanoasperity for three graphene-coated colloidal probes. The area of the topographically-highest nanoasperity is roughly estimated by processing the topographies in (a),(c),(e) with the "flooding" option of the WSXM software. This method is based on setting all the image values below a given threshold value $h_{th}$ (selected by the user) to a constant value. The resulting image resembles a picture of a flooded region of land (Figure S6, right column). Once the process is finished WSXM can calculate the area $A_{isl}(h_{th})$ of the "drought islands". The threshold value $h_{th}$ was varied for each topography in small steps of $0.1 - 0.2$nm below the maximum surface height $h_{max}$, to find out the threshold $h_{th}^*$ separating the regime of single-asperity contact ($0 < h_{th} \leq h_{th}^*$, only one island in the "flooded topography") from a multi-asperity contact ($h_{th} > h_{th}^*$, two or more islands in the "flooded" topography). The area of the topographically-highest nanoasperity was estimated as $A_{isl}(h_{max} - h_{th}^*)$. We conventionally assumed that $h_{th}^*$ values that are more than 1nm apart from $h_{max}$ (i.e. $h_{max} - h_{th}^* > 1$nm) are unphysical. In fact the penetration depth of a nanoasperity into graphite is expected to be of only $\delta \sim 0.3 - 0.6$nm at the normal load $F_N = 100$nN (from Hertzian contact theory $\delta = (F_N^2/E^2 R)^{1/3}$ with the effective Young modulus $E \sim 30$GPa and a tentative curvature radius $R = 50 - 300$nm for the nanoasperity). In such specific situations the area of the topographically-highest nanoasperity was assumed to be $A_{isl}(h_{max} - 1$nm$)$. We obtained that $A_{isl}$ is $\sim 2 \times 10^2$ nm$^2$ for all the beads. Arrows in the original AFM topographies (left column) and "flooded topographies" (right column) highlight the location of the single "drought island".

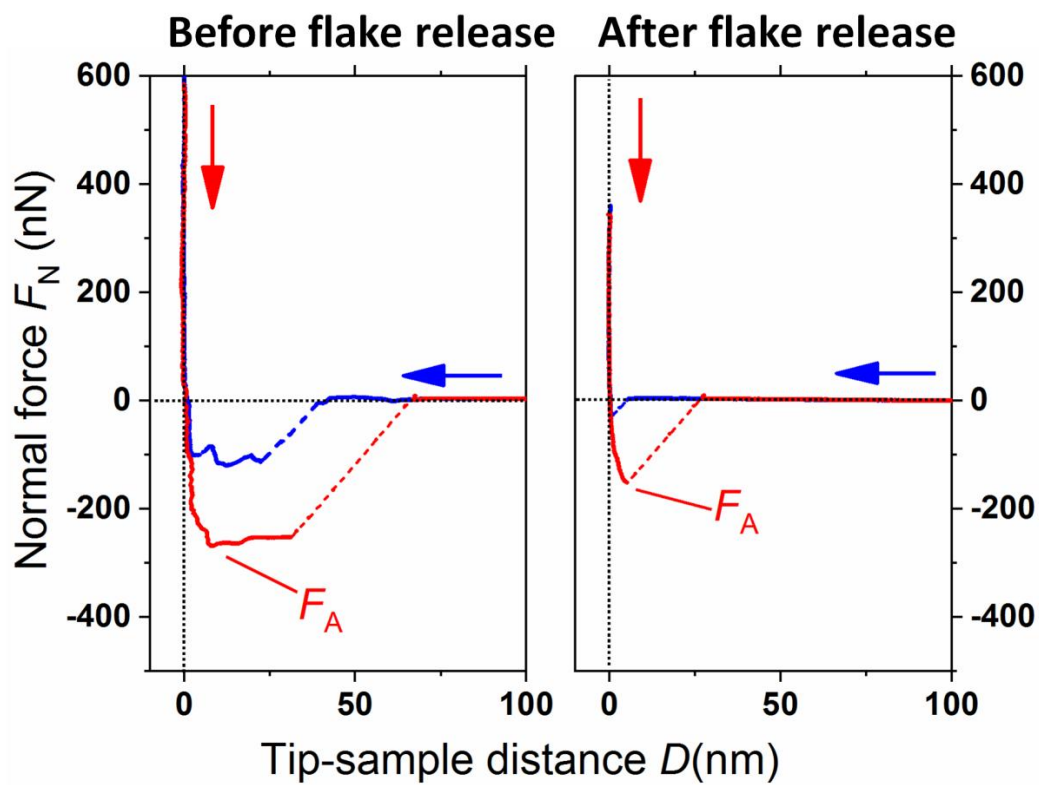

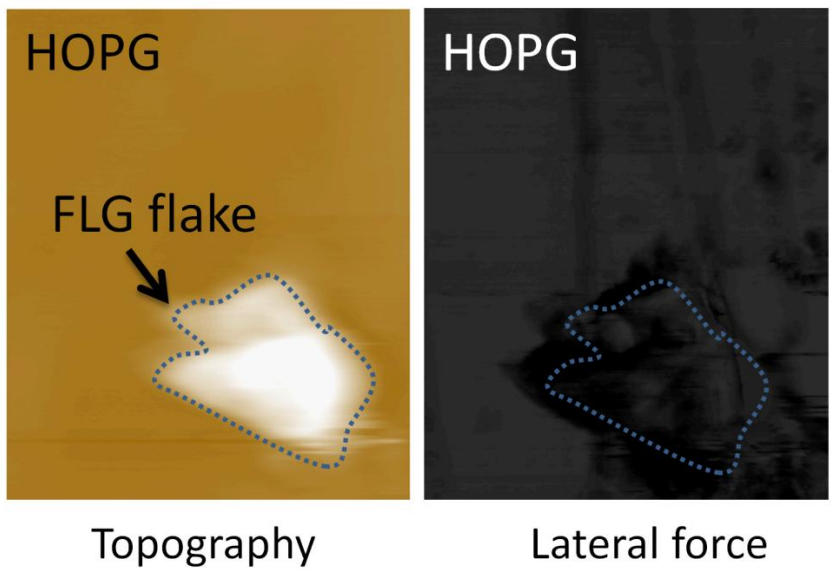

**Figure S7.** Variation of the long-ranged behavior of two spectroscopic curves (top) acquired with the very same graphene-coated colloidal probe on HOPG. Variation is triggered by the release of a loosely attached FLG flake from the coated probe to HOPG (occurred while sliding on HOPG in between the two spectroscopic experiments).

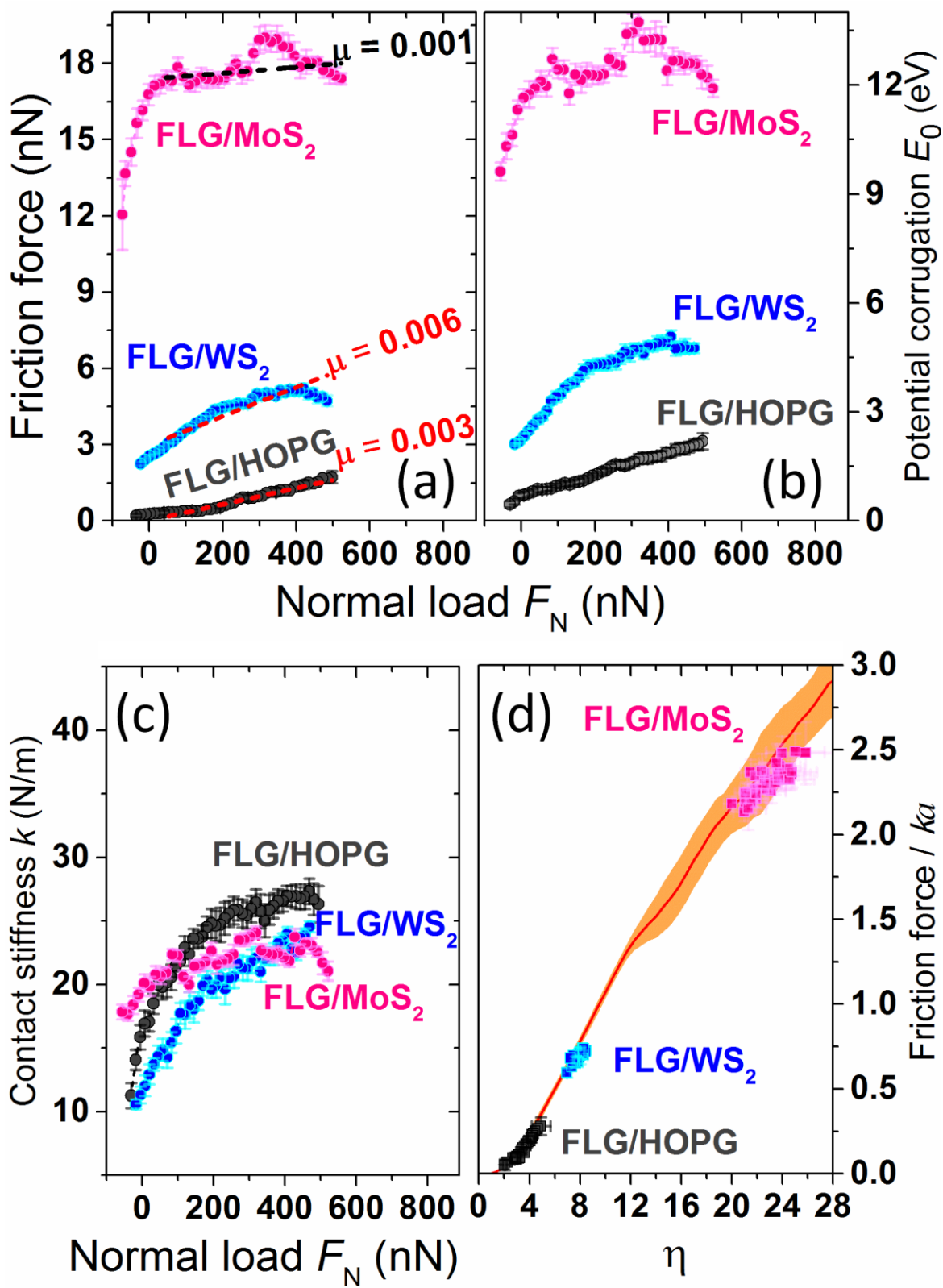

**Figure S8.** (a) Representative $F_f$ vs $F_N$ characteristics measured by means of the colloidal probe shown in Figure S3a and Figure S5 (see also main text Figure 3c,d) for the three layered sliding junctions FLG/HOPG, FLG/WS$_2$ and FLG/MoS$_2$ respectively ($v = 33$ nm/s). (b)-(c) Load-dependent variation of the interfacial parameters $E_0$ and $k$. (d) Comparison of experimental data with the PT model. Data are represented by square symbols as in Figure 6i.

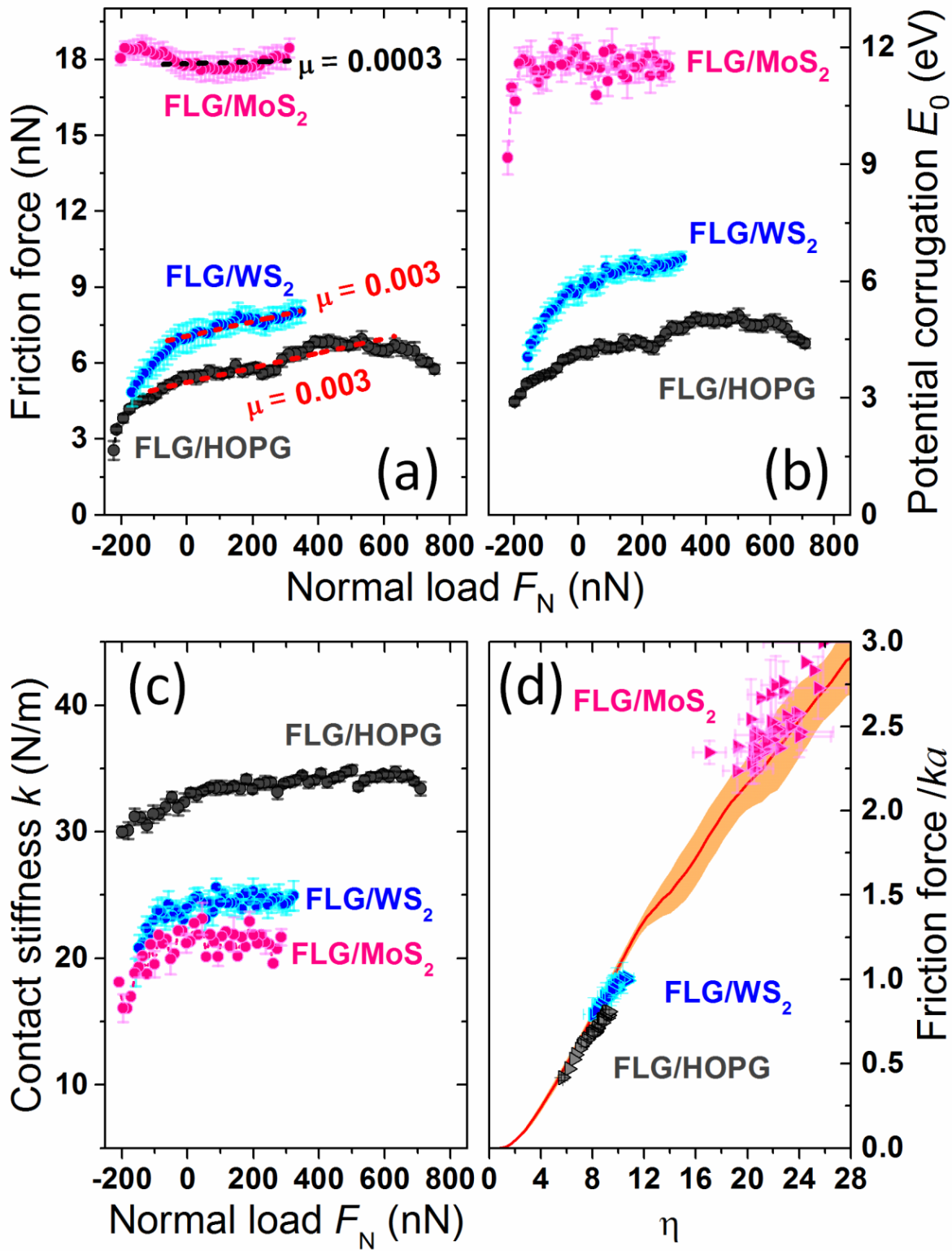

**Figure S9.** (a) Representative $F_f$ vs $F_N$ characteristics measured for the three layered sliding junctions FLG/HOPG, FLG/WS$_2$ and FLG/MoS$_2$ respectively ($v = 33$nm/s). An AFM micrograph of the colloidal probe used in this case is reported in Figure S3e. (b)-(c) Load-dependent variation of the interfacial parameters $E_0$ and $k$. (d) Comparison of experimental data with the PT model. Data are represented by triangles, as in Figure 6i.